\documentclass[11pt]{article}
\usepackage{vmargin}
\setpapersize{USletter}           
\setmargrb{1.4in}{0.5in}{1.4in}{1in}

\usepackage{latexsym}
\usepackage{graphicx}
\usepackage{amssymb,amsmath}
\usepackage{epstopdf}
\usepackage{footnote}

\usepackage{stmaryrd}

\usepackage{sectsty}
\sectionfont{\normalsize}
\subsectionfont{\normalsize}
\subsubsectionfont{\normalsize}

\def\be{\begin{equation}}
\def\ee{\end{equation}}
\def\ba{\begin{array}}
\def\ea{\end{array}}
\def\e{{\rm e}}
\def\ljump{\mbox{$[ \! [$}}
\def\rjump{\mbox{$] \! ]$}}

\begin{document}



\begin{center}
{\bf Linearized Theory of Traffic Flow}\\
Tal~Cohen\footnote{Corresponding author. talco@mit.edu} and Rohan Abeyaratne\\
Department of Mechanical Engineering\\
Massachusetts Institute of Technology\\
Cambridge, MA 02139, USA\\[2ex]
\end{center}

\begin{center}
{\bf Abstract}
\end{center}

The equation of motion of a general class of macroscopic traffic flow models is linearized around a steady uniform flow. A closed-form solution of a boundary-initial value problem is obtained, and it is used to describe several phenomena.  Specifically, the scenarios examined involve a smooth velocity field in stop-and-go traffic, a discontinuous velocity field with shock waves in a traffic light problem, and discontinuous displacement fields in a problem where a single platoon of vehicles splits into two, and later merges back into one.


\section{Introduction.}\label{intro}

In modern industrial metropolitan areas,  traffic congestion has become an increasing inconvenience with both social and economic implications.  The design of more efficient and safe vehicular transportation systems requires the ability to simulate traffic behavior at both the macroscopic and microscopic levels. Appropriate models should be able to capture a wide range of known traffic phenomena at minimal computation times and with as few model parameters as possible.

New sensor, communications, automation and other technologies are being rapidly developed, and in turn, leading to improved transportation systems that require new models in order to be simulated. For example, traditional traffic models assume that a driver is aware only of the traffic conditions ahead of his/her vehicle, whereas sensors are now frequently installed on both the front and rear bumpers of vehicles which makes it possible for the driver (or an autonomous vehicle control system) to take into account the traffic conditions behind it as well, e.g. \cite{Horn}. New traffic models continue to be developed as transportation systems evolve.

Comprehensive reviews of the theories of traffic flow can be found in \cite{Heibing2001, Hoog, helbing2007, Treiber}.
Some traffic models are deterministic while others are statistical, e.g. \cite{Alperovich2008}.  ``Microscopic'' (discrete) models track each vehicle, e.g. \cite{Chandler, Newell}, while ``macroscopic'' (continuum) models treat the vehicles in a smeared-out manner and so study the collective behavior of a stream of vehicles, e.g. \cite{LW2, Richards, Payne1971, AwR, Zhang}. As described in Section \ref{20140801.somemodels}, the most commonly used continuum models involve a system of nonlinear partial differential equations, often hyperbolic, that exhibit a variety of nonlinear wave phenomena.  Solutions are usually determined numerically since the equations are too complex to solve analytically. The numerical solutions can often be very sensitive to the computational details, and one must be careful when interpreting results.

While certain traffic phenomena are inherently nonlinear, others are not. Often one is concerned with perturbations of a steady uniform flow so that even a nonlinear model can be linearized about some motion. Linearized models can often be solved analytically, and therefore serve as a useful tool in probing the behavior of traffic. In the present paper we examine such a linearization. Needless to say we are not arguing that traffic models should be linear.  Instead our point is that linearized models can often provide useful insight, which when needed may be followed-up with a numerical solution of the full nonlinear model.

Specifically, in this paper we look at traffic flows that are perturbations from a steady uniform flow. We linearize a general class of equations of motion and solve a boundary-initial value problem in closed form. Then, using that solution, we examine three specific scenarios. What varies from one scenario to the next is the smoothness of the boundary condition (and therefore the solution). In the first, the lead vehicle changes its speed continuously. In the second it changes discontinuously. And in the third, an interior vehicle in a platoon changes its speed such that the platoon splits into two, and so the displacement field itself becomes discontinuous.

We  observe a certain parallel between the phenomena described in these three problems and in the mathematical modeling of the mechanics of an elastic body. Depending on the boundary conditions (i.e. the nature of the loading), the strain and velocity fields in the body will automatically be either continuous or involve shocks.  On the other hand the displacement field in the body can also become discontinuous (in the event of fracture), but now elasticity theory by itself will not generate such solutions. One must input information from outside of elasticity, about the conditions for fracture initiation and propagation, just as we shall see is needed in a traffic problem where a platoon splits into two.

Some physical theories lend themselves to an Eulerian formulation (e.g. fluid mechanics) while other theories are simplest when cast in a Lagrangian framework (e.g. solid mechanics). Eulerian formulations have been the framework of choice in traffic modeling, in part, because this field can trace its origin to its analogy to fluid flows.  A Lagrangian setting is convenient in certain circumstance, e.g when one needs to consider events at specific vehicles such say a breakdown, or in ensuring that information does not propagate ahead of the vehicles. In some recent studies of traveling periodic traffic waves the authors found a Lagrangian formulation to be more convenient, \cite{Greenberg, Seibold}.  Microscopic models are effectively Lagrangian since they track individual vehicles.  Of course an Eulerian framework has its own advantages, and in any case one formulation can be mapped into the other so one is not intrinsically correct or incorrect. Here we shall adopt a Lagrangian framework throughout.

This paper is organized as follows: in the next section we make some further remarks on the Lagrangian framework, briefly describe microscopic models, review the development of several standard macroscopic traffic models,  and describe some phenomena observed in traffic flows. In Section \ref{20140803.sec-form} we formulate the equations of motion and jump conditions, and also remark on a linear (as against linearized) theory of traffic flow.  Then in Section \ref{20140802.secstab} we discuss the stability of the model. We also illustrate our formulation by specializing it to the Aw-Rascle-Zhang model  \cite{AwR, Zhang}.  We state and solve the boundary-initial value problem in closed form in Section \ref{20140818.sec4}. In Section \ref{20140803.sec-someprobs} we examine three specific problems concerning stop-and-go waves, a traffic light problem, and the separation of a stream of traffic into two and their subsequent merging.


\section{Background.} \label{20180819.background}


\subsection{Lagrangian framework.} 

The earliest model of traffic flow was based on its analogy with fluid flow, and just as in fluid mechanics, the modeling was done within an Eulerian framework, Lighthill and Whitham \cite{LW2}, Richards \cite{Richards}. The primary field variables were the traffic density $\rho$ and vehicle velocity $v$, and these were taken to be functions of the current location $y$ of a vehicle  and time $t$. Thus the theory was developed for the fields $\rho(y,t)$ and $v(y,t)$ (one of which can be replaced by the traffic flux $q = \rho v$). Most generalizations of this first model have continued to be presented within this framework.

In a Lagrangian framework one identifies a vehicle by its location $x$ in some reference configuration. It is convenient to take the primary fields of the theory to be the headway $\lambda$ and the vehicle velocity $v$, and to consider them as functions of $x$  and $t$: $\lambda(x,t), v(x,t)$. The headway is related to the reciprocal of the traffic density and so it reflects {\it the distance between vehicles}. In this paper we shall use a Lagrangian formulation throughout. Of course any Lagrangian formulation can be transformed into an Eulerian formulation and vice versa by making use of the mapping $y(x,t)$.


\subsection{Microscopic (discrete) models of traffic flow.}

We briefly mention here the ``microscopic'' modeling of traffic, commonly referred to as car-following models. Following Chandler et al. \cite{Chandler}   and Newell \cite{Newell},  one considers $N+1$ vehicles moving along the $x$-axis. The position of the $n$th vehicle at time $t$ is $y_n(t)$ and its velocity is $v_n(t) = \dot{y}_n(t)$. 
The distance between vehicles is of greater interest than the location of each vehicle and so the theory is formulated in terms of the headway $\lambda_n(t) = y_{n-1}(t) - y_n(t)$, i.e. the distance between the $n$th vehicle and the vehicle immediately ahead of it, and the velocity $v_n$.

Suppose, for example, that the driver of a vehicle controls its acceleration based on $(i)$ the rate at which its distance from  the vehicle immediately in front of it is changing, $(ii)$ this distance itself, and $(iii)$ its own velocity: 
$$
\dot{v}_n(t) = a_n(\dot{\lambda}_n(t),  \lambda_n(t), {v}_n(t)).
$$
The acceleration functions $a_n$ here are to be determined empirically, and depend on the vehicles, roadway, road conditions, types of traffic etc.  Under uniform conditions $a_n$ will be independent of $n$. Particular choices of $a_n$ correspond to particular car-following models, several examples of which can be found in Chapters 10 and 11 of Treiber and Kesting \cite{Treiber}.

When the density of traffic is sufficiently high, one expects to be able to take the continuum limit of a microscopic model thus deriving a partial differential equation based model that characterizes a related macroscopic model.


\subsection{Macroscopic (continuum) models of traffic flow.}\label{20140801.somemodels}

Here we briefly review the development of some continuum traffic models. Our presentation will be Lagrangian even though these models are usually described in the literature in (equivalent) Eulerian form.

The continuum theory is formulated in terms of the headway $\lambda(x,t)$ and the vehicle velocity $v(x,t)$, that  are the continuum counterparts of $\lambda_n(t)$ and $v_n(t)$ and will be defined more carefully in Section  \ref{20140808.secpde}.
The traffic model comprises of an equation of motion, e.g.
$$
{v}_t = a({\lambda}_t,  \lambda_x, \lambda, {v}),
$$
together with the compatibility requirement $\lambda_t = v_x$, where the subscripts $x$ and $t$ denote partial differentiation. The acceleration function $a$ is determined empirically.

Different choices of $a$ correspond to different traffic models. 
The earliest and simplest model of traffic flow is due to Lighthill and Whitham \cite{LW2} and Richards \cite{Richards} where the differential equation above reduces to an algebraic equation
\be
v = V(\lambda)  \label{20140801.LWR}
\ee
with $V$ being an empirical function with $V'(\lambda) >0$. Despite its simplicity the LWR model (\ref{20140801.LWR})   is able to capture many phenomena observed in traffic flow, e.g. see Chapter 8 of Treiber and Kesting \cite{Treiber}.

The density and flux of traffic are $\rho = \rho_0/\lambda$ and $q = \rho v$ respectively where $\rho_0$ is the traffic density in the reference configuration.  One can construct the flux function $Q(\rho) = \rho V(\rho_0/\rho)$. The function $Q$, or a graph of it, is called the ``{\it fundamental diagram}'' and appears frequently in the literature. Given $Q$ one can find $V$ and vice versa.

The primary limitation of the LWR model is that the velocity at some instant $t$ is given by the headway at that same instant $t$ since, when written out fully, (\ref{20140801.LWR}) reads $v(x,t) = V(\lambda(x,t))$.  In the terminology of continuum mechanics, the model has no history dependence. One way to remedy this is to include time derivatives of the fields in the model, the simplest example of which is
\be
v_t = \frac{V(\lambda) - v}{\tau}. \label{20140801.LWRtimelag}
\ee
The empirical function $V(\lambda)$ characterizes the behavior of the system in steady state and so is called the {\it equilibrium velocity function}.

The main limitation of the preceding model is the fact that the response of some vehicle (vehicle $x$) depends only on the local conditions at that vehicle.  In the terminology of continuum mechanics, it describes a local theory. The simplest way in which to account for the fact that the driver is aware of the vehicles ahead of it is to include appropriate spatial gradient terms of the basic fields in the model, as for example in 
\be
v_t = \frac{V(\lambda) - v}{\tau} + \frac{\nu_0}{\tau} \frac{\lambda_x}{\lambda^2},
 \label{20140801.LWRtimelaganti}
\ee
Whitham \cite{GBW-Book}.  The term involving $\lambda_x$ is sometimes referred to as the ``anticipation''.
The Payne-Whitham (PW) model (Payne \cite{Payne1971}) generalizes  (\ref{20140801.LWRtimelaganti}) as
\be
v_t = \frac{V(\lambda) - v}{\tau} - p'(\lambda) \lambda_x, \label{20140801.PW}
\ee
where $p(\lambda)$ is an empirical function with $p'(\lambda)<0$.

One of the limitations of the PW model is that, under certain conditions, it predicts vehicles to move backwards, Daganzo \cite{Daganzo}.  Also, one of the (Lagrangian) characteristic speeds of the model (\ref{20140801.PW}) is positive and so it allows information to propagate ahead of the vehicles. In order to avoid these limitations Aw and Rascle \cite{AwR} and Zhang \cite{Zhang}, proposed the ARZ model
\be
v_t = \frac{V(\lambda) - v}{\tau} - h'(\lambda) \lambda_t, \label{20140801.AR}
\ee
where $h(\lambda)$ is an empirical function with $h'(\lambda)<0$. (Note that $\lambda_t = v_x$ and so this model does account for an awareness of other vehicles.)

Observe that if the traffic flow is both steady, so that time derivatives vanish, and uniform, so that spatial derivatives vanish, then all of the preceding models reduce to the LWR model  (\ref{20140801.LWR}).

The preceding models of traffic flow lead to systems of hyperbolic partial differential equations and will therefore, under certain conditions, exhibit solutions that involve shocks. Some authors add a term involving $v_{xx}$ to the model in order to regularize the theory, thus replacing the shocks by regions of rapid (but smooth) variation, as, for example, in
the Kerner-Konh\"{a}user (KK) model \cite{KK}
\be
v_t = \frac{V(\lambda) - v}{\tau} + \frac{\nu_0}{\tau} \frac{\lambda_x}{\lambda^2} + \mu_0 \left[\frac{v_{xx}}{\lambda} - \frac{\lambda_x v_x}{\lambda^2}\right].   \label{20140801.KK}
\ee
See also, e.g. Ge and Han \cite{GeHan} and Zhang \cite{Zhang2003}.  However the physical significance of such terms is unclear, and therefore it is difficult to know what the appropriate functional form for such terms should be.

Different models have different strengths and weaknesses, new models continue to be developed and enhanced. Existing models, including the LWR, PW and ARZ models, continue to be used extensively. Recent reviews of traffic models can be found, for example, in Hoogendoorn and Bovy \cite{Hoog}, Treiber and Kesting \cite{Treiber}.


\subsection{Stop-and-go waves.}

A ``stop-and-go wave'' generally refers to a propagating perturbation that causes each vehicle to decrease and increase its velocity several times. These waves are triggered by the delay in a driver's response, especially, according to Laval and Leclercq \cite{stopgo}, by aggressive or timid driving. An intuitive example of such a wave is in the sudden deceleration of a particular vehicle in relatively dense traffic. Under such conditions, due to the lag in the following driver's response, that driver must decelerate even more rapidly to avoid collision. The driver will overshoot the new desired speed and will then have to accelerate. This can happen a few times before the vehicle settles into its new desired speed and headway. Conceivably, the third driver will have to respond even more abruptly, and so on. This chain of events evolves into a traffic wave. 

Depending on the conditions of the road, as this wave propagates, it can either disperse or intensify, possibly even leading to the complete stopping of some downstream vehicle.  The stability of the traffic flow plays an important role here. Relatively dense traffic flows can become unstable and evolve into an intensifying stop-and-go mode of flow. Such stop-and-go behavior, even if it does not lead to vehicles having to stop, may lead to unsafe driving conditions and increased fuel consumption.    
The origin of such a wave can be a small perturbation in the motion of one vehicle, and yet, the unstable traffic behavior can evolve into a local jam. Such a jam is frequently referred to as a \textit{phantom jam}, since it appears to occur for no particular reason.

According to the literature survey by Sch{\"o}nhof and Helbing \cite{helbing2007}, (Eulerian) stop-and-go waves propagate against the  direction of the vehicle flow  at a speed of $15\pm5~[km/h]$.
A comprehensive description of stop-and-go waves and phenomena observed in traffic flow can be found in \cite{Heibing2001, Hoog, helbing2007, Treiber},


\section{Formulation.}\label{20140803.sec-form}

\subsection{Governing differential equations.} \label{20140808.secpde}

Consider a platoon of vehicles moving along a one-lane highway -- the $x$-axis. In a Lagrangian formulation a vehicle is identified by its position\footnote{Alternatively one can let $x$ denote the vehicle number which is simply a nondimensional real number. This leads to a dimensional headway $\lambda = y_x$. We choose to stay with the more conventional approach used in continuum mechanics. The two approaches are of course trivially related through a convenient length scale associated with the reference configuration.}  $x$ in a  reference configuration. Suppose that in the reference configuration the platoon is associated with the interval $-L \leq x \leq 0$.  During a traffic flow the vehicle $x$ is located at $y(x,t)$ at time $t$.
The (nondimensional) headway $\lambda(x,t)$ and vehicle velocity $v(x,t)$ are defined by
\be
\lambda = y_x, \qquad v = y_t , \label{20140803.lamv}
\ee
where the subscripts $x$ and $t$ denote partial differentiation.  In the terminology of continuum mechanics, $\lambda$ would be the stretch.
The macroscopic model of traffic flow that we consider is
\be
\left.
\ba{lll}
{v}_t &=& a({\lambda}_t,  \lambda, {v}),\\[2ex]

{\lambda}_t &=& v_x,\\
\ea\right\} , \qquad -L \leq x \leq 0, \ t \geq 0. \label{20140805.genmodel}
\ee
Equation (\ref{20140805.genmodel})$_2$ expresses compatibility between (\ref{20140803.lamv})$_1$ and (\ref{20140803.lamv})$_2$. The equation of motion (\ref{20140805.genmodel})$_1$ is a statement of the fact that a driver adjusts the acceleration of his vehicle based on the rate at which the distance to the vehicle ahead of it is changing, the distance to the vehicle ahead of it, and its own velocity. The acceleration function $a$ is assumed to be known empirically.

First consider a steady uniform motion
\be
y(x,t) = \lambda_0 x + v_0 t. \label{20140805.sum}
\ee
Substituting (\ref{20140805.sum}) into  (\ref{20140803.lamv}), (\ref{20140805.genmodel})  leads to the algebraic equation
\be
a(0, \lambda_0, v_0) = 0. \label{20140805.azero}
\ee
Thus the headway $\lambda_0$ and velocity $v_0$ in a steady uniform motion are not independent but must be related by (\ref{20140805.azero}).  Suppose that (\ref{20140805.azero}) can be solved for $v_0$ to give
\be
v_0 = V(\lambda_0).  \label{20140805.lamv}
\ee
This implies that the function $V$ must obey
\be
a(0, \lambda, V(\lambda) ) = 0 \label{20140805.azerotwo}
\ee
for all headways $\lambda$.  Since $V$ is the function that relates the headway to the velocity in a steady uniform motion, it is in fact the {\it equilibrium velocity function} introduced previously below (\ref{20140801.LWRtimelag}).

Next consider a motion close to the steady uniform motion (\ref{20140805.sum}):
\be
y(x,t) = \lambda_0 x + v_0 t + u(x,t)  \label{20140805.nearby}
\ee
where $\lambda_0$ and $v_0$ are related by (\ref{20140805.lamv}), (\ref{20140805.azerotwo}), and $u$ is {\it the displacement (departure)  from the steady uniform motion} and is suitably small. Substituting (\ref{20140805.nearby}) into  (\ref{20140803.lamv}), (\ref{20140805.genmodel}) and linearizing the result leads to
\be
u_{tt} = a_1 u_{xt} + a_2 u_x + a_3 u_t \label{20140805.govpde}
\ee
where we have used (\ref{20140805.azero}) and set
\be
a_1 = \left. \frac{\partial a}{\partial \lambda_t} \right |_{(\lambda_{t}, \lambda, v) = (0, \lambda_0, v_0)},
\ 
a_2 = \left. \frac{\partial a}{\partial \lambda} \right |_{(\lambda_{t}, \lambda, v) = (0, \lambda_0, v_0)},
\
a_3 = \left. \frac{\partial a}{\partial v} \right |_{(\lambda_{t}, \lambda, v) = (0, \lambda_0, v_0)}.\label{20140805.aaa}
\ee
Observe by differentiating (\ref{20140805.azerotwo}) with respect to $\lambda$ and evaluating the result at $\lambda_0$ that
\be
a_2 + a_3 V'(\lambda_0) = 0. \label{20140805.a2a3}
\ee
Note from (\ref{20140805.aaa}) and (\ref{20140805.lamv}) that the coefficients $a_k$ are functions of $\lambda_0$.

Equation (\ref{20140805.govpde}) is hyperbolic (provided $a_1 \neq 0$) and can be written 
in the form of a generic second-order wave equation
  \be
\tau  \left(\frac{\partial}{\partial t} + c_- \frac{\partial}{\partial x}\right)\left(\frac{\partial u}{\partial t} + c_+ \frac{\partial u}{\partial x}\right) + \left( \frac{\partial u}{\partial t} + c_0 \frac{\partial u}{\partial x}\right) = 0 \label{20140731.genwaveeqn}
 \ee
by setting $c_+ = - a_1$, $c_- = 0$, $c_0 = a_2/a_3$ and $\tau = -1/a_3$.
Here $c_+$ and $c_-$ are the wave speeds (characteristic speeds) of this second-order equation and $\tau$ is a time constant (that, as we will see later, is a related to the lag in a driver's response). If the second-order terms were absent, (\ref{20140731.genwaveeqn}) is a first-order wave equation with wave speed $c_0$.  
This motivates us to introduce the notation
\be
c = a_1, \qquad \tau = - 1/a_3, \qquad c_0 = - a_2/a_3 = V'(\lambda_0), \label{20140709.7}
\ee
so that $-c$ and $-c_0$ denote the speeds of the second-order and first-order waves respectively. In obtaining the second expression for $c_0$ above we have used (\ref{20140805.a2a3}). Observe that the relative importance of the second-order terms, as compared with the first-order terms, is determined by $\tau$.  In terms of the wave speeds and time constant, the governing differential equation takes the final form
\be
\tau ({{u}}_{tt} - c {{u}}_{xt})  +  {{u}}_t - c_0 {{u}}_x = 0.
\label{20140709.8}
\ee

This model involves three quantities: $c$ -- the propagation speed (against the flow of traffic) of the second-order waves, which therefore is the speed with which the first signal propagates; $c_0$ -- the propagation speed (against the traffic flow) of the first-order waves which by given through the equilibrium velocity function as $c_0 = V'(\lambda_0)$, 
and $\tau$ -- the lag time in a driver's response. Note from (\ref{20140805.aaa}), (\ref{20140805.a2a3}) and (\ref{20140709.7}) that $c$ and $c_0$ depend on the headway $\lambda_0$ of the steady uniform flow (as well as on the parameters of the model contained within the acceleration function $a(\lambda_t, \lambda, v)$). An example will be given in Section \ref{20140805.secex}.


\subsection{Jump conditions.}

The first derivatives $u_x$ and $u_t$ of a solution of the hyperbolic equation (\ref{20140709.8}) may have discontinuities. Suppose that $u(x,t)$ is continuous everywhere on the domain of interest of $x,t$-plane, and that it varies  smoothly except that $u_x$ and $u_t$ have finite jump discontinuities across a curve\footnote{Since the equation (\ref{20140709.8})  is linear, these discontinuities are not, strictly speaking, shocks in the sense in which this term is used in the mathematical literature. For simplicity of terminology we will refer to them as shocks.}  $x=s(t)$. Then, by writing the governing equation (\ref{20140709.8}) in the form of a pair of conservation laws we conclude that the jump conditions
\be
\ljump u_t \rjump \dot{s} + c \ljump  u_t \rjump = 0, \qquad \ljump u_x \rjump \dot{s} + \ljump  u_t \rjump = 0, \label{20140731.jumpsxyz}
\ee
must hold at a shock. Here, for any field $g(x,t)$, we write $\ljump g \rjump = g(s(t)+, t) - g(s(t)-, t)$ for the jump in $g$.   One way in which to satisfy  (\ref{20140731.jumpsxyz}) is to have 
\be
\dot{s} = 0, \qquad \ljump u_t \rjump = 0.
\ee
 A shock with (Lagrangian) speed $\dot{s} = 0$  does not propagate with respect to the vehicles (and is referred to as a ``contact discontinuity''). For $\dot{s} \neq 0$, (\ref{20140731.jumpsxyz}) is equivalent to 
\be
\dot{s} = - c ,  \qquad  \ljump u_t \rjump = c \ljump u_x \rjump. \label{20140731.reducedjumps}
\ee
Note that the two possible shock speeds $0$ and $-c$ are identical to the two characteristic speeds, which, of course, is expected since the system is linear.


\subsection{Linear model.}

In the special case when $a(\lambda_t, \lambda,v)$ is a linear function of its arguments, one obtains the equation of motion (\ref{20140709.8}) {\it exactly}, without linearization. This is not as drastic an assumption as it might first seem, since, suppose the fundamental diagram is triangular (as is often assumed, see Chapter 8 of Treiber and Kesting \cite{Treiber}).  The corresponding congested flow branch of the flux function $Q(\rho)$ is 
$$
Q(\rho) = q_* \frac{\rho_j - \rho}{\rho_j - \rho_*}, \qquad \rho_* < \rho < \rho_j,
$$
where $\rho$ is the density of traffic, $\rho_j$ is the jamming density when the vehicles are stopped bumper-to-bumper and there is zero traffic flux, and $\rho_*$ is the density at which the flux has its maximum value $q_*$. On using the flux-density relation $q = \rho v$ and the density-headway relation $\lambda = 1/\rho$ we find the corresponding equilibrium velocity function for congested flow to be the following linear function:
$$
V(\lambda) = \frac{q_* \rho_j}{\rho_j - \rho_*} \left(   \lambda -  1/\rho_j\right), \qquad 1/\rho_j < \lambda <  1/\rho_*.
$$
 Note in particular that the first-order wave speed $c_0 = V'(\lambda_0)$ now has the illuminating expression
 $$
 c_0 = \frac{q_* \rho_j}{\rho_j - \rho_*} ,
 $$
in terms of the maximum traffic flux, the traffic density when the flux is maximum, and the jamming density. The results in this paper would be exact solutions in this linear theory provided one ensures that the headway $\lambda(x,t)$ is everywhere in the range $1/\rho_j < \lambda <  1/\rho_*$.


\section{String stability and an example.}\label{20140802.secstab}

The governing differential equation (\ref{20140709.8}) involves three quantities: the wave speed $c$, the wave speed $c_0 = V'(\lambda_0)$ of the associated first-order wave equation, and the driver's lag time $\tau$. 
We expect information to not travel ahead of the vehicles, i.e. we expect (Lagrangian) waves to not propagate in the positive $x$-direction.  If they did, a downstream vehicle would be aware of an upstream vehicle. Thus we shall require the wave speeds $-c$ and $-c_0$ to be negative. Also, the lag time in a driver's response cannot be negative and so $\tau$ should be positive. Thus we require
\be
c > 0, \qquad c_0 > 0, \qquad \tau > 0. \label{apriori}
\ee

Given an explicit traffic model (such as the ARZ model considered below in Section \ref{20140805.secex} one can determine expressions for $c$ and $c_0$ (and perhaps $\tau$ in some cases) in terms of the headway $\lambda_0$ of the underlying steady uniform flow. For the ARZ model we will see that both wave speeds decrease as $\lambda_0$ increases (traffic gets less dense) while the driver's lag time remains constant. Since these two quantities are functions of the headway $\lambda_0$, and they are {\it not}  ``parameters'' of the model in the usual sense. The example in Section \ref{20140805.secex} will illustrate this.

\subsection{String stability.}

In this section we determine the restrictions  imposed on the quantities $c, c_0$ and $\tau$ by stability considerations. There are several notions of stability in the traffic flow literature as discussed by Treiber and Kesting in Chapter 15 of  \cite{Treiber}. It is sufficient in this study to limit attention to {\it string stability}.

To examine the string stability of the steady uniform motion we consider a perturbation in the (usual) (Fourier component) form 
\be
u(x,t) = \e^{i(kx+\omega t)} , \qquad x <0, \ t >0, \label{sc1} 
\ee
where $\omega$ is real but $k$ may be complex. Thus each vehicle $x$ undergoes a time harmonic oscillations with frequency $\omega$. The reciprocal of the real part of $k$ is the wave length of the spatial modulation, the imaginary part of $k$ characterizes the growth/decay of the amplitude of oscillation with $x$.  String stability requires the amplitude of oscillation to decay as $|x|$ increases, which will happen if and only if the imaginary part of $k$ is negative.

Substituting (\ref{sc1}) into (\ref{20140709.8})  leads to an algebraic equation that is linear in $k$. Solving it for $k$ yields
$$
k =   \frac{\omega  + i \omega^2 \tau}{ c_0 +i\omega \tau c} .
$$
The condition for the imaginary  part of $k$ to be negative is
\begin{equation}
\tau (c - c_0) > 0. \label{20140730.locstringstab}
\end{equation}
Keeping (\ref{apriori}) in mind, this requires that 
\be
c > c_0.
\ee
Thus {\it stability/instability} reduces to whether the {\it ratio of the wave speeds} $c/c_0$ is greater than or less than unity. In what follows we will consider traffic flow under both stable and unstable conditions. 

Perhaps it is worth noting that since $c$ and $c_0$ both depend on the headway $\lambda_0$ the stability inequality $c(\lambda_0) > c_0(\lambda_0)$ yields, in general, the ranges of headway for which the steady uniform flow is stable.

\subsection{Example.}\label{20140805.secex}

For purposes of illustration, we now determine the wave speeds $c$ and $c_0$ for the ARZ model (\ref{20140801.AR}): from (\ref{20140805.lamv}), (\ref{20140805.aaa}) and (\ref{20140709.7}) we find that they are
\be
c = - h'(\lambda_0), \qquad c_0 = V'(\lambda_0) .\label{20140805.paramarz}
\ee
The parameter $\tau$ in the ARZ model (\ref{20140801.AR}) is related to the lag time in a driver's response (which depends on the driver's reaction time as well as  the driver's ``temperament'', \cite{helbing2007}). The quantity $\tau$ in the present study is readily shown to coincide with the parameter $\tau$ in the ARZ model and so we shall refer to it as the {\it driver's lag time}.

Suppose that the specific functions $V$ and $h$ in the ARZ model are
\be
V(\lambda) = v_{max} \left( 1 - \frac{\lambda_{min}}{\lambda} \right), \qquad h(\lambda) = h_0 \ln \left( \frac {\lambda_{min}}{\lambda} \right) \ , \label{20140814.specialarz}
\ee
where the positive constants $v_{max}, \lambda_{min}$ and $h_0$ are model parameters. Observe that the value of this equilibrium velocity function $V$ increases monotonically from $0$ to $v_{max}$ as the headway increases from $\lambda_{min}$ to infinity. For the choice (\ref{20140814.specialarz}), the wave speeds (\ref{20140805.paramarz}) specialize to
\be
c = \frac{h_0}{\lambda_0}, \qquad c_0 = \frac{v_{max} \lambda_{min}}{\lambda^2_0}.
\ee
The wave speeds $c$ and $c_0$ depend on the material parameters and the headway $\lambda_0$.
Note that the inequalities $c > 0$ and $c_0 > 0$ are automatic. The requirement for stability, i.e.  $c/c_0 > 1$, can now be written equivalently as
\be
\lambda > \lambda_c \qquad {\rm where} \quad \lambda_c = \frac{v_{max} \lambda_{min}}{h_0}.
\ee
Thus the steady uniform flow is stable if the headway is large ($\lambda > \lambda_c$), i.e. for sufficiently light traffic, while it is unstable for small values of headway ($\lambda < \lambda_c$), i.e. for sufficiently dense traffic.

It is important to mention that for certain other choices of $V$ and $h$ in the ARZ model, the regime of unstable flow corresponds to {\it intermediate traffic densities}, both low and high densities being stable.


\section{A boundary-initial value problem.} \label{20140818.sec4}

Consider a semi-infinite platoon of vehicles $- \infty < x \leq 0$. Suppose that for all times prior to the initial instant,  the vehicles undergo a steady uniform motion: $y(x,t) = \lambda_0 x + v_0 t$ for $t < 0$.  Now suppose that at time $t=0$ the leading vehicle $x=0$ changes its velocity so that from then on  
$u_t(0,t) = v_f(t)$ for $t > 0$ where $v_f$ is given. 
Since the vehicles were undergoing a steady uniform motion for $t<0$, we may assume that at the initial instant
$u(x, 0) = 0, u_t(x,0) = 0$ for $x < 0$,
recalling that $u$ denotes the displacement from the steady uniform motion.
Thus in summary, given $v_f(t)$, we wish to solve the following boundary-initial value problem for $u(x,t)$:
\be\left.
\ba{lcll}
\tau( {u}_{tt} - c\, {{u}}_{xt}) + {{u}}_t - c_0 {{u}}_x  = 0 
, \qquad & x < 0, \ t> 0,\\[2ex]

{u}(x,0) = 0, \qquad {u}_t(x,0) = 0 \qquad &  x<0, \\[2ex]

{u}(0,t) = {{u}}_f(t), \qquad &t > 0,\\
\ea\right\} \label{20140709.22}
\ee
where 
\be
u_f(t) = \int_0^t v_f(\xi) \, d\xi \label{20140802.bc}
\ee
is the (known) displacement of the leading vehicle.


\subsection{Analytical solution of boundary-initial value problem.} \label{20140803.sec-analsoln}

We now use Laplace transforms to solve the problem (\ref{20140709.22}) using the notation $\mathcal{L}$ and $\mathcal{L}^{-1}$ to denote the forward and inverse transform operators.   Let $\overline{u}(x,s) = \mathcal{L}[u(x,t)]$ and $\overline{u}_f(s) = \mathcal{L}[u_f(t)]$. Taking the transform of
 the partial differential equation (\ref{20140709.22})$_1$ leads to the ordinary differential equation
  \be
( c_0 + c \tau s) \overline{u}_x  - (\tau s^2 + s) \overline{u} = 0, \label{20140709.25}
 \ee
 and the boundary condition (\ref{20140709.22})$_3$ similarly yields
\be
 \overline{u}(0,s)  =  \overline{u}_f(s) .  \label{20140709.23xx}
\ee
The initial conditions (\ref{20140709.22})$_2$ have been used in obtaining (\ref{20140709.25}).

The solution of (\ref{20140709.25}) with (\ref{20140709.23xx}) is
\be
 \overline{u}(x,s) = \overline{u}_f(s) \,  \e^{\phi(s) x} , \label{20140709.28}
\ee
where
\be
\phi(s) =  \frac{s}{c \tau} \, \left( \frac{\tau s + 1}{ s + \vartheta } \right)  =  \frac{s}{c}  - \frac{\varphi}{c} + \frac{\beta}{s + \vartheta} \ , \label{20140709.31}
\ee 
and we have set
\be
 \vartheta = \frac{c_0}{c \tau} \, >0, \qquad 
 \varphi = \frac{1}{\tau} \left(\frac{c_0}{c} - 1\right) , \qquad 
 \beta =  \frac{c_0}{c^2 \tau^2} \left( \frac{c_0}{c}  - 1 \right) . \label{20140709.32}
\ee
Observe that in the stable case, $c_0/c <1$, both $\varphi$ and $\beta$ are negative. However since we wish to examine both stable and unstable flows we shall not restrict the sign of $\varphi$ and $\beta$.
The solution $u(x,t)$ is given by the inverse transform of (\ref{20140709.28}):
\be
u(x,t) = \mathcal{L}^{-1} \left[ \overline{u}_f(s) \, \e^{\phi(s)x}\right] = \int_0^t u_f(\xi) p(t-\xi) \, d\xi \, ,
\label{20140813.5}
\ee
where we have  set
\be
p(t) =  \mathcal{L}^{-1} \left[ \e^{\phi(s)x}\right] \, , \label{20140813.6}
\ee
and used the standard convolution theorem for the inverse Laplace transform of the product of two function.
We now proceed to find $p(t)$.

Recall that, for any function $f(t)$ that has a Laplace transform $\overline{f}(s)$, the following are well-known identities:
\begin{equation}\left.
\ba{lll}
\mathcal{L}^{-1}[e^{\alpha s}
\bar f(s)]=\mathcal{H}(t+\alpha)f(t+\alpha), \\[2ex]

\mathcal{L}^{-1}[\overline{f}(s-\alpha)] = \e^{\alpha t} f(t),\\[2ex]

\mathcal{L}^{-1}[\overline{f}(s/\alpha)] = \alpha \ f(t\alpha), \qquad \alpha>0, \\
\ea
\right\}
\label{20140813.threeintoone}
\end{equation}
where $\mathcal{H}(t)$ is the Heaviside (step) function and $\alpha$ is a real constant.
Also, from Laplace transform tables one finds that
\begin{equation}
\mathcal{L}^{-1}\left[ \e^{1/s} \right]= \frac{1}{\sqrt{t}} I_1( \sqrt{4t}) + \delta(t), \qquad
\mathcal{L}^{-1}\left[ \e^{-1/s} \right]= \frac{1}{\sqrt{t}} J_1( \sqrt{4t}) + \delta(t),
\label{20140813.4}
\ee
where $I_1(x) $ is a modified Bessel function of the first kind, $J_1(x) $ is a Bessel function of the first kind and $\delta(t)$ is the delta function.

In order to find $p(t)$, we start by combining (\ref{20140709.31}) with (\ref{20140813.6}) and then using (\ref{20140813.threeintoone})$_1$ to write
\begin{equation}
p(t) = \mathcal{L}^{-1}[ e^{\phi(s)x }]
= e^{- {x \varphi}/ {c}} \ 
 \mathcal{L}^{-1} \left[ e^{ {sx}/{c} }e^{{\beta x}/{(s+\vartheta)}} \right] =
  e^{- {x \varphi}/ {c}} \mathcal{H}(t+x/c) q(t+x/c), \label{20140813.7}
\end{equation}
where 
\begin{equation}
q(t) =   \mathcal{L}^{-1} \left[ e^{{\beta x}/{(s+\vartheta)}} \right] =  \e^{-\vartheta t}\mathcal{L}^{-1} \left[ e^{{\beta x}/{s}} \right] , \label{20140813.9}
\end{equation}
and we have used (\ref{20140813.threeintoone})$_2$ in obtaining the second equality.
Finally, by using (\ref{20140813.threeintoone})$_3$ we have
\begin{equation}
\mathcal{L}^{-1} \left[ e^{{\beta x}/{s}} \right] = 
\left\{
\ba{rll}
\beta x \, r_1(\beta x t), \qquad & \beta < 0,\\[2ex]

|\beta x| \, r_2(|\beta x t|), \qquad & \beta > 0,\\
\ea\right. \label{20140813.10}
\end{equation}
where $r_1(t)$ and $r_2(t)$ are the respective functions given by the right hand sides of (\ref{20140813.4})$_1$ and (\ref{20140813.4})$_2$.

We can now use (\ref{20140813.10}) in (\ref{20140813.9}) to find $q(t)$, and then use the result in (\ref{20140813.7}) to find $p(t)$. Substituting this expression for $p$ into (\ref{20140813.5}) yields the following explicit expression for {\it the solution} of the boundary-initial value problem:
\be
u(x,t) = \mathcal{H}(t+x/c) \left[
\e^{-\frac{x\varphi}{c}}  u_f(t+x/c) + \e^{-\frac{x\varphi}{c}} \int_{-x/c}^t u_f(\xi+x/c) \Phi(x, t-\xi) d\xi \right]
\label{20140814.16}
\ee
where $\mathcal{H}$ is the Heaviside function and
$\Phi$ is defined by
\begin{equation}
\Phi(x,t)=e^{-\vartheta t}\left\{
\ba{rll}
\sqrt{\frac{\beta x}{t} }I_1( \sqrt{4\beta x t})\qquad &\text{for}\quad \beta<0,  \\[2ex]

-\sqrt{\frac{\left|\beta x \right|}{t}} J_1\left( \sqrt{ \left| 4\beta x t \right| }\right)
 \qquad &\text{for} \quad \beta>0.  \\
 \ea \right. \label{20140814.flowfnc}
\end{equation}
We refer to $\Phi$ as  the \textit{traffic flow function}.
The vehicle velocity and headway can now be found from (\ref{20140803.lamv}) and (\ref{20140805.nearby}):
\be
v(x,t) = v_0 + u_t(x,t), \qquad \lambda(x,t) = \lambda_0 + u_x(x,t). \label{20140815.vlam}
\ee

The spatial and temporal evolution of the displacement $u(x,t)$ is characterized by (\ref{20140814.16}) which obeys the equation of motion as well as the initial and boundary conditions.  As expected, the first signal moves backwards with (Lagrangian) velocity $c$. The displacement field is undisturbed prior to the arrival of this wave: $u(x,t) = 0$ for $t+x/c < 0, x < 0$. The intensity of the wave depends on the disturbance $u_f(t)$ caused by the leading vehicle.

The first term within the square brackets  in (\ref{20140814.16})  describes a straightforward propagation of the boundary disturbance but with its amplitude varying exponentially with $|x|$. In the stable regime we have $c_0/c < 1$, and therefore $\varphi < 0$, and so (recalling that $x<0$) the exponential term describes a decay of the solution. In the unstable case it describes exponential growth.


\section{Results and Discussion.}\label{20140803.sec-someprobs}

In order to gain some insight into the character of the general solution (\ref{20140814.16}) we first discuss its asymptotic behavior. Thereafter three specific boundary-initial value problems will be presented and discussed.


\subsection{Asymptotic behavior of solution.}

We know from Section \ref{20140803.sec-analsoln} that the solution of the boundary-initial value problem (\ref{20140709.22}) is given by the inverse Laplace transform of $ \overline{u}_f(s) \, \e^{\phi(s) x}$ where $\phi(s)$ is given by (\ref{20140709.31}). Following Whitham \cite{GBW-Book}, 
the behavior of $u$ near the leading wave front $x+c t = 0$ can be obtained by first replacing $\phi(s)$ by its
leading approximation $\phi(s) \sim s/c + (1-c_0/c) (1/(c\tau))$ as $s \to \infty$ and then carrying out the inverse transformation.
This leads to
\be
u(x,t) \sim u_f(t+x/c) \, \exp \left [  - \left( 1 - \frac{c_0}{c} \right) \, \frac{|x|}{c \tau} \right] , \label{20140817.asy}
\ee
 which is appropriate near leading wave front $x+c t = 0$. It describes a wave that damps out on the length scale $c \tau$ as it propagates at the speed $-c$. As $\tau$ decreases (and the first-order equation grows in importance) this term becomes smaller (at every $x$). Note that (\ref{20140817.asy}) is precisely the first term in the square brackets in (\ref{20140814.16}).


\subsection{Stop-and-go waves.} \label{chap:perturbation}

Next we discuss the exact solution (\ref{20140814.16}) for a specific choice of the boundary perturbation $v_f(t)$ by evaluating the integral numerically. The specific example in this section involves a continuous function $v_f(t)$. We shall examine the solution for various values of $\tau$ and $c_0/c$.  Recalling for the particular ARZ model introduced Section \ref{20140805.secex} that
$c_0/c = \lambda_c/\lambda_0$,
where $\lambda_c$ is the critical value of headway at which stability is lost; thus varying $c/c_0$ is equivalent to varying $\lambda_0/\lambda_c$.  Our results are not limited to this ARZ model.

Suppose that at time $t=0$ the lead vehicle begins to gradually decreases its velocity to $v_0 - v_*$ and then gradually increases it back to its original value $v_0$ over a time interval $T$. From then on it maintains its velocity at $v_0$.  To describe this we take 
\begin{equation}
v_{f}(t)=
\begin{cases}
-v_* \sin(\pi t/T), & 0 \leq t \leq T, \\
~~~0, & t \geq T. \\
\end{cases} 
\end{equation} 
From this, (\ref{20140805.nearby}), (\ref{20140802.bc}) and (\ref{20140814.16})  we calculated the current position $y(x,t)$ of a vehicle $x$ at time $t$.  The trajectories of the vehicles -- $y(x,t)$ versus $t$ at constant $x$ -- were then plotted. Likewise, using (\ref{20140815.vlam})$_1$ we determined the velocity field $v(x,t)$.

\begin{figure}[h]
\begin{centering}
\centerline{\includegraphics[width=1\textwidth]{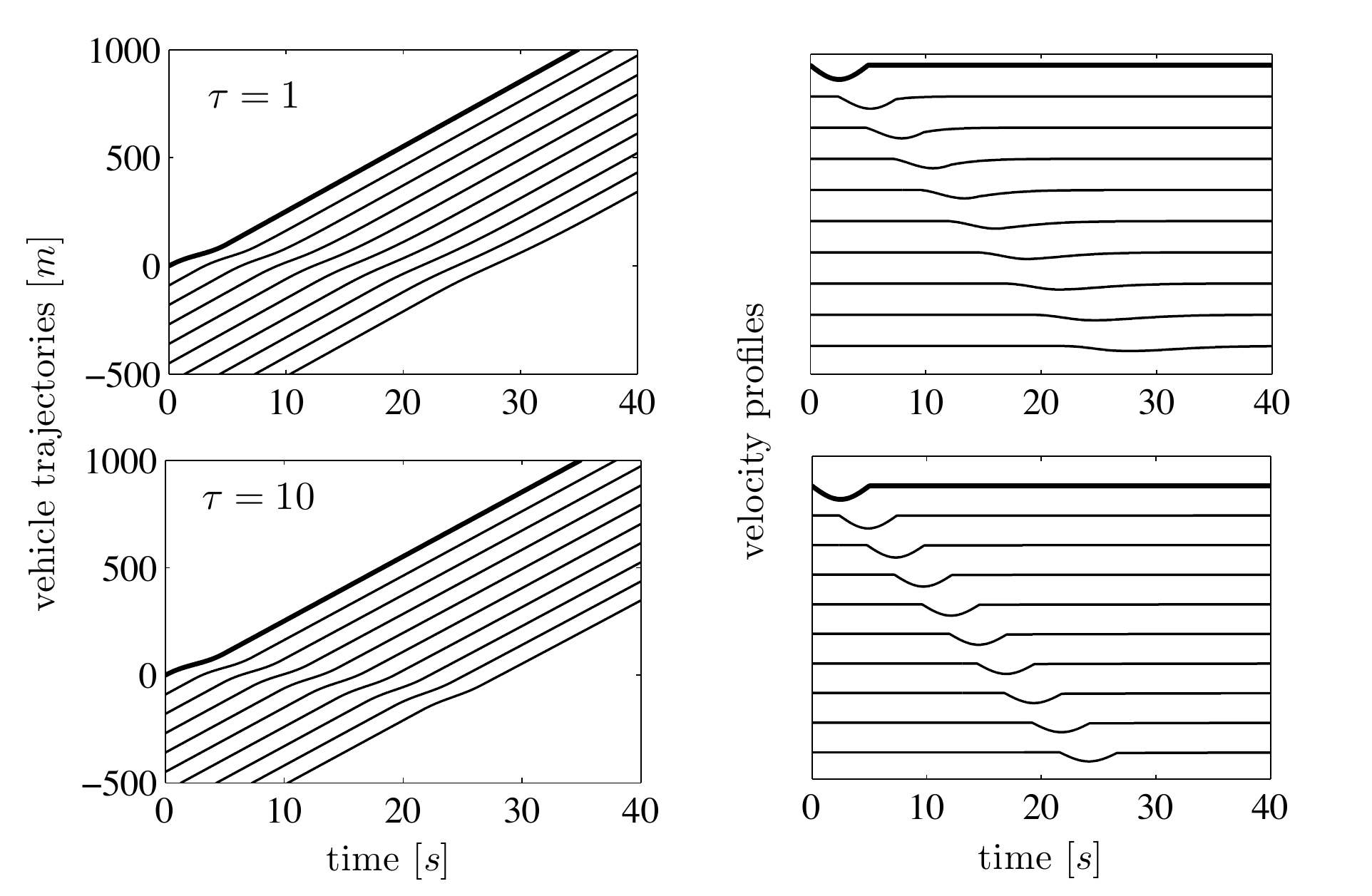}}
 \end{centering}
\caption{Traffic wave under stable conditions: $c/c_0 = 1.25$. Response of a traffic platoon to a pulse perturbation in the leader's velocity. The panels on the left show the trajectories of individual vehicles. The panels on the right show the corresponding velocity histories; each curve is associated with a different vehicle. The thick black curve corresponds to the leading vehicle. The figures have been plotted with $c_0=30 [m/s]$ and $v_*=0.3 v_0$. The top row of figures is for $\tau=1$, the bottom row for $\tau=10$.}
\label{perturbation1}
\end{figure}

The first series of calculations were carried out under stable conditions. Specifically, we took $c/c_0 =1.25$ and considered two values of $\tau$, viz. $\tau =1.0$ and $10.0$. 
The results are shown in Figure \ref{perturbation1}. The panels on the right show plots of vehicle velocity versus time for different vehicles $x$. The bold curve at the top in each panel corresponds to the lead vehicle and $|x|$ increases as one moves down a panel. The panels on the left show the associated vehicle trajectories. 
For both values of $\tau$ the perturbation travels upstream at the speed $c$,  decreases in magnitude as it propagates, and disperses. Each vehicle essentially mimics the behavior of the lead vehicle, slowing down and then speeding up, but by progressively decreasing amounts.  The time that it takes for the perturbation to dissipate is sensitive to the quantity $\tau$. The perturbation persists for a longer time the larger $\tau$ is. This is expected since larger values of $\tau$ correspond to drivers who respond more slowly.

\begin{figure}[h]
\begin{centering}
\centerline{\includegraphics[width=1\textwidth]{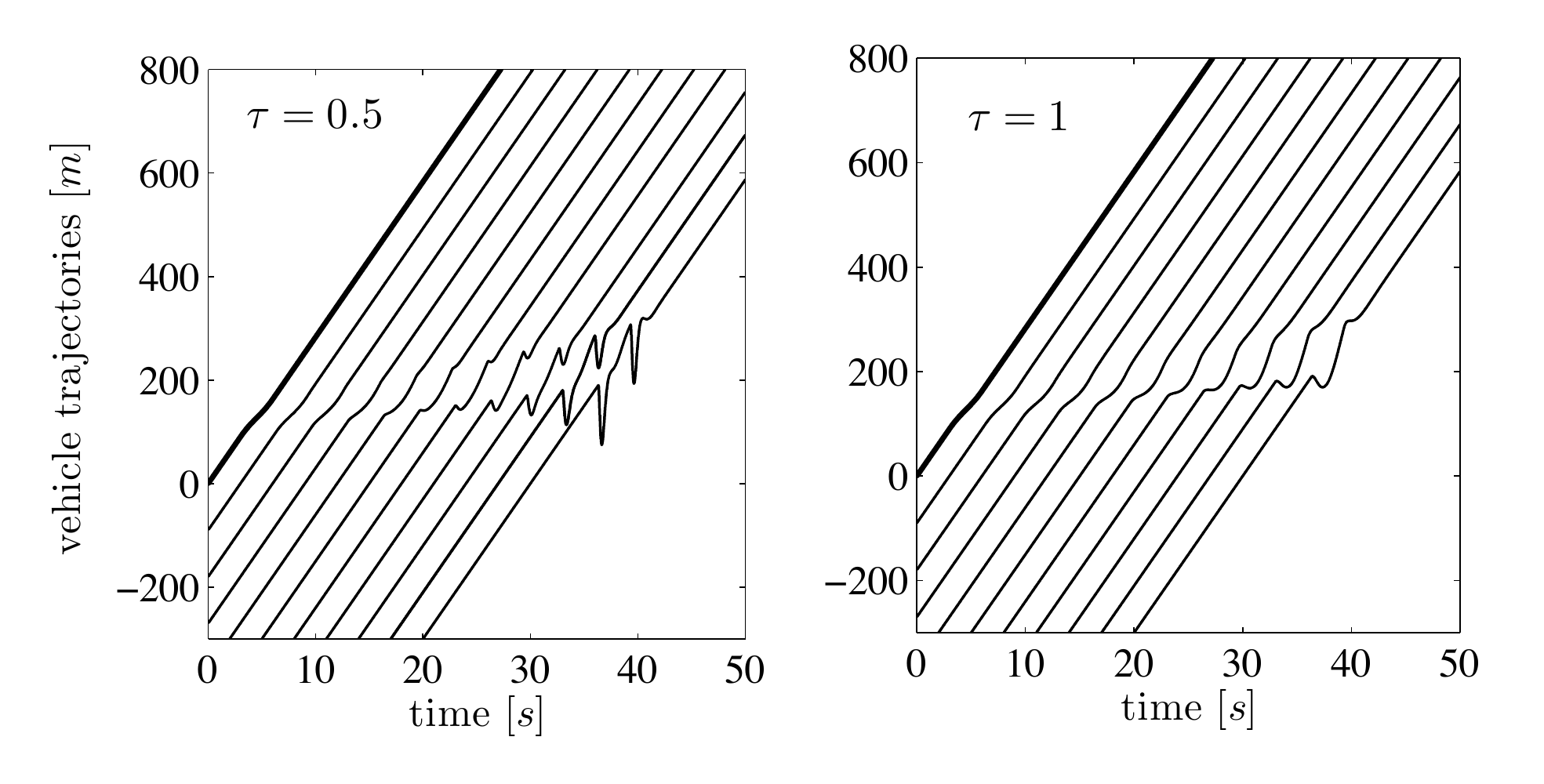}}
 \end{centering}
\caption{Development of stop-and-go waves under unstable conditions: $c/c_0 = 0.9$. Response of a traffic platoon to a pulse perturbation in the leader's velocity. The panels show the trajectories of individual vehicles. The thick black curves correspond to the leading vehicle. The figures have been plotted with $c_0=30 [m/s]$ and $v_* = 0.3 v_0$; for the figure on the left $\tau=1.0$, on the right $\tau = 0.5$.}
\label{perturbation2}
\end{figure}

The second series of calculations were carried out under unstable conditions, specifically with  $c/c_0=0.90$. The results are displayed in Figure \ref{perturbation2} which shows  how the disturbance intensifies as it propagates and how the single pulse evolves into a wave with multiple oscillations. This is a classical stop-and-go wave where vehicles accelerate and decelerate several times.  This behavior can be explained intuitively as being a consequence of aggresive driving. Once the driver who is following notices the deceleration of the vehicle in front, s/he responds excessively thus creating a large gap between it and the vehicle ahead.  To return to the original configuration it is then necessary to accelerate. A second follower must respond even more abruptly to the deceleration of the vehicle in front, and while the vehicle in front re-accelerates the second follower must close an even larger gap. Eventually the solution leads to the intersection of vehicle trajectories, after which time the solution is no longer valid.

   
\subsection{Traffic light problem.} \label{chap:shock}

We now consider a problem where the velocity of the lead vehicle (the boundary condition) changes discontinuously .  

Suppose, for example, that the (entire) platoon of vehicles is stopped at a red light for $t<0$. Suppose that the light turns green and the leading vehicle ($x=0$) abruptly accelerates, increasing its velocity instantaneously to some value $v_0 + v_*$. It then maintains this velocity for a time $T$, at which point it reaches a second red light and stops abruptly. Thus the boundary condition is 
\be
v_f(t) = 
\left\{
\ba{lll}
v_*, \qquad & 0 < t < T,\\[2ex]

0, \qquad & t > T.\\
\ea\right. \label{20140818.traficlightbvp}
\ee

Discontinuous changes in velocity are not physically realistic (except perhaps in the event of an accident), and in order to model this behavior, the mathematical solution will involve shocks. 
Each time the lead vehicle changes its velocity abruptly, it nucleates a shock which then propagates upstream at the (Lagrangian) speed $c$. The headway and velocity suffer jump discontinuities across each shock, and they are related by (\ref{20140731.reducedjumps}). The strength of the shock (the magnitude of the discontinuity) decays exponentially as it propagates according to (\ref{20140817.asy}).

The solution to this problem was obtained from (\ref{20140814.flowfnc}) and (\ref{20140818.traficlightbvp}) and some results are shown in Figure \ref{Abrupt}. The panel on the right shows plots of vehicle velocity versus time for different vehicles $x$. The bold curve at the top corresponds to the lead vehicle and $|x|$ increases as one moves down that panel. The panel on the left shows the associated vehicle trajectories.

We note that there is a shock in our solution, both at the sudden start and the sudden stop. According to nonlinear theories that include an entropy condition, a shock would be involved only at the sudden stop, with a continuous variation of the fields (a fan) at a sudden start. In this regard, we observe that the shocks in the linearized theory decay exponentially, and it appears from Figure \ref{Abrupt} that this decay is faster at the shock triggered by the sudden start, so that the flow very quickly becomes (more or less) continuous; the shock strength decays more slowly at the shock nucleated by the sudden stop.

 \begin{figure}[h]
\begin{centering}
\centerline{\includegraphics[width=1\textwidth]{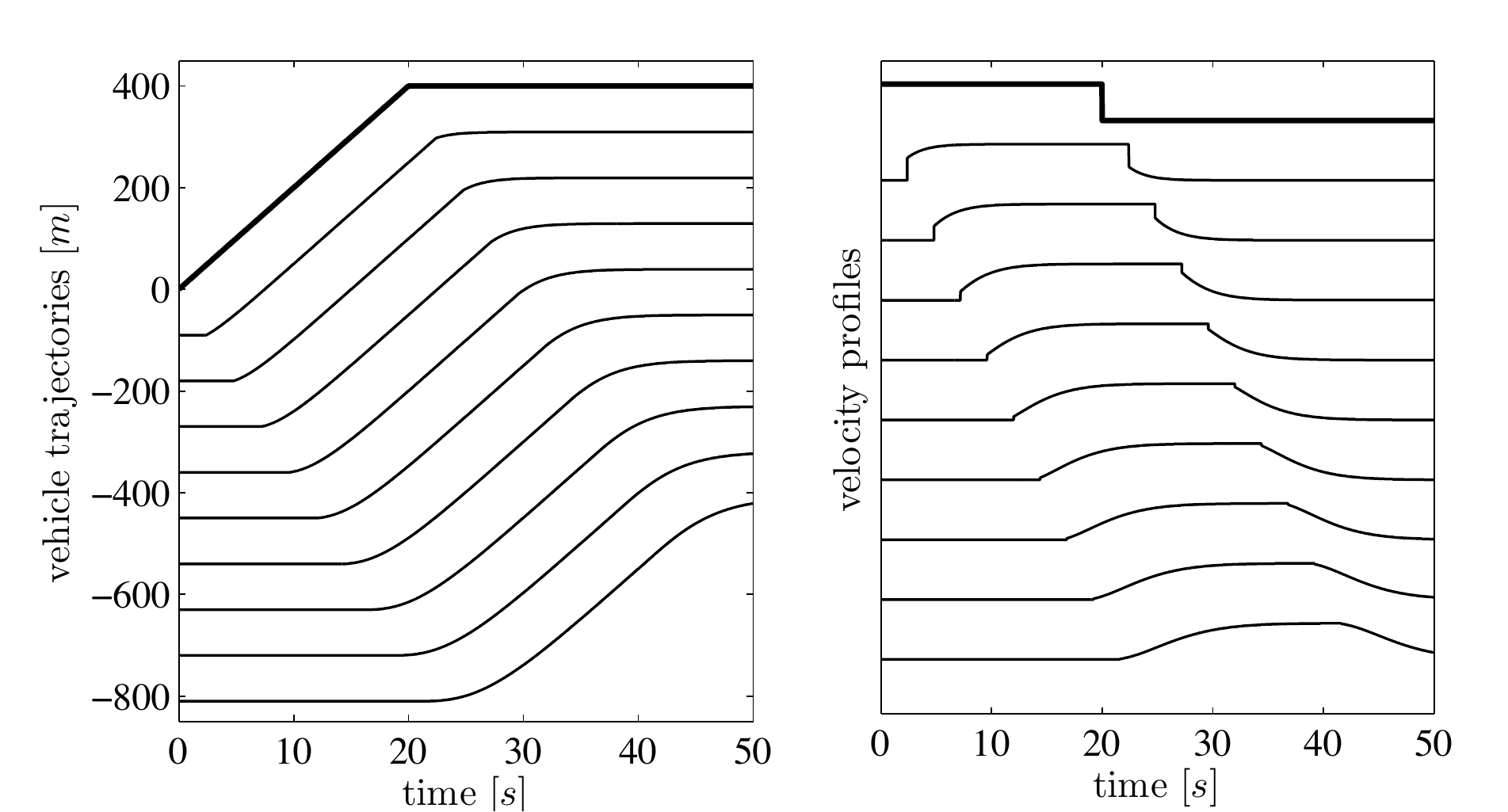}}
 \end{centering}
\caption{Traffic light change. All vehicles are initially stationary at a red light. When the light turns green, the lead vehicle abruptly increases its velocity and then maintains it. When it reaches a second red light, it abruptly stops. The left panel shows the vehicle trajectories with the corresponding velocity profiles shown on the right hand panel. The thick black lines correspond to the lead vehicle.  The figure has been drawn with $c_0=30 [m/s],~~c/c_0 =1.25$ and $ \tau=1$. }
\label{Abrupt}
\end{figure}


\subsection{Disturbance within the flow. Separating and merging of traffic.}

We now consider a problem in which the displacement field itself becomes discontinuous.

Consider again a platoon of vehicles $x<0$ undergoing a steady uniform flow. At some instant $t_1 > 0$ the vehicle $x_1$ in the interior of the platoon decides to adjust its speed (perhaps in response to a traffic light it sees up ahead). The vehicles ahead of it, $x_1 < x < 0$, continue to travel as before. The perturbation in the motion of vehicle $x_1$ will of course affect the motion of all vehicles behind it, $x < x_1$.

From time $t_1$ onwards we have two platoons of vehicles, and the vehicle $x_1$ is now the lead vehicle for the rear platoon. Its motion {\it must therefore be prescribed externally}. Suppose that the perturbation in the velocity of vehicle $x_1$ is $v_1(t) = {\dot{u}}_1(t)$ for $t_1 < t < t_2$. Then we have the following boundary and initial conditions, where as before, $u(x,t)$ is the displacement from the steady uniform motion:
\be
u(x, t_1) = 0, \qquad u_t(x, t_1) = 0, \qquad x < x_1
\ee
\be
u_t(x_1, t) = {v}_1(t), \qquad  t_1 < t < t_2.
\ee
Clearly the response of the vehicles is given by the prior solution (\ref{20140814.16}) provided we shift the origin of space, time from $(0,0)$ to $(x_1, t_1)$ and replace $u_f$ by $u_1$. In particular, a wave will be nucleated at $(x_1, t_1)$ and propagate backward with speed $c$. The trajectories associated with such a scenario in shown in Figure \ref{complicated}.

Note the singular nature of the flow at $(x_1, t_1)$.  A single vehicle trajectory splits into two. The last vehicle of the platoon in front is $x_1^{+}$, the lead vehicle of the platoon behind is $x_1^{-}$. The trajectory of the vehicle $x_1^{+}$ remains undisturbed and is given by $y(x_1^{+},t) = \lambda_0 x_1 + v_0 t$. The trajectory of the vehicle $x_1^{-}$ is $y(x_1^{-},t) = \lambda_0 x_1 + v_0 t + u_1(t)$ for $t_1 < t < t_2$.  In the terminology of fluid mechanics there is ``flow separation''.  For continuity we must have $u_1(t_1) = 0$. To avoid collision, we must have $u_1(t) < 0$ for $t_1 < t < t_2$. Observe that the region of the $x,t$-plane between these two trajectories is a ``vacuum'' -- there are no vehicles in it.

Suppose that at some later time $t_2 (> t_1)$ the vehicle $x_1^{-}$ catches up with the vehicle $x_1^{+}$ and that from then on the two platoons travel as a single platoon once more. Thus $u_1(t_2) = 0$.  Moreover, from then on,  this vehicle is no longer the lead vehicle and so its motion cannot be prescribed externally but must be found via the differential equations.  If ${v}_1(t_2) \neq 0$ the velocity of this vehicle will change discontinuously at the instant $t_2$ when it merges with the platoon in front, thus generating a shock wave that will propagate upstream from $(x_1, t_2)$.  Smooth merging requires ${v}_1(t_2) = 0$. This is reminiscent of the Kutta condition in aerodynamics that is imposed when there is a flow around a sharp corner.

Figure \ref{complicated} shows the vehicle trajectories for a specific scenario illustrating the phenomena described above.  In the calculations underlying the figure, the vehicle $x_1^-$ was taken to first decelerate, and then accelerate, both linearly.

\begin{figure}[h]
\begin{centering}
\centerline{\includegraphics[width=0.8\textwidth]{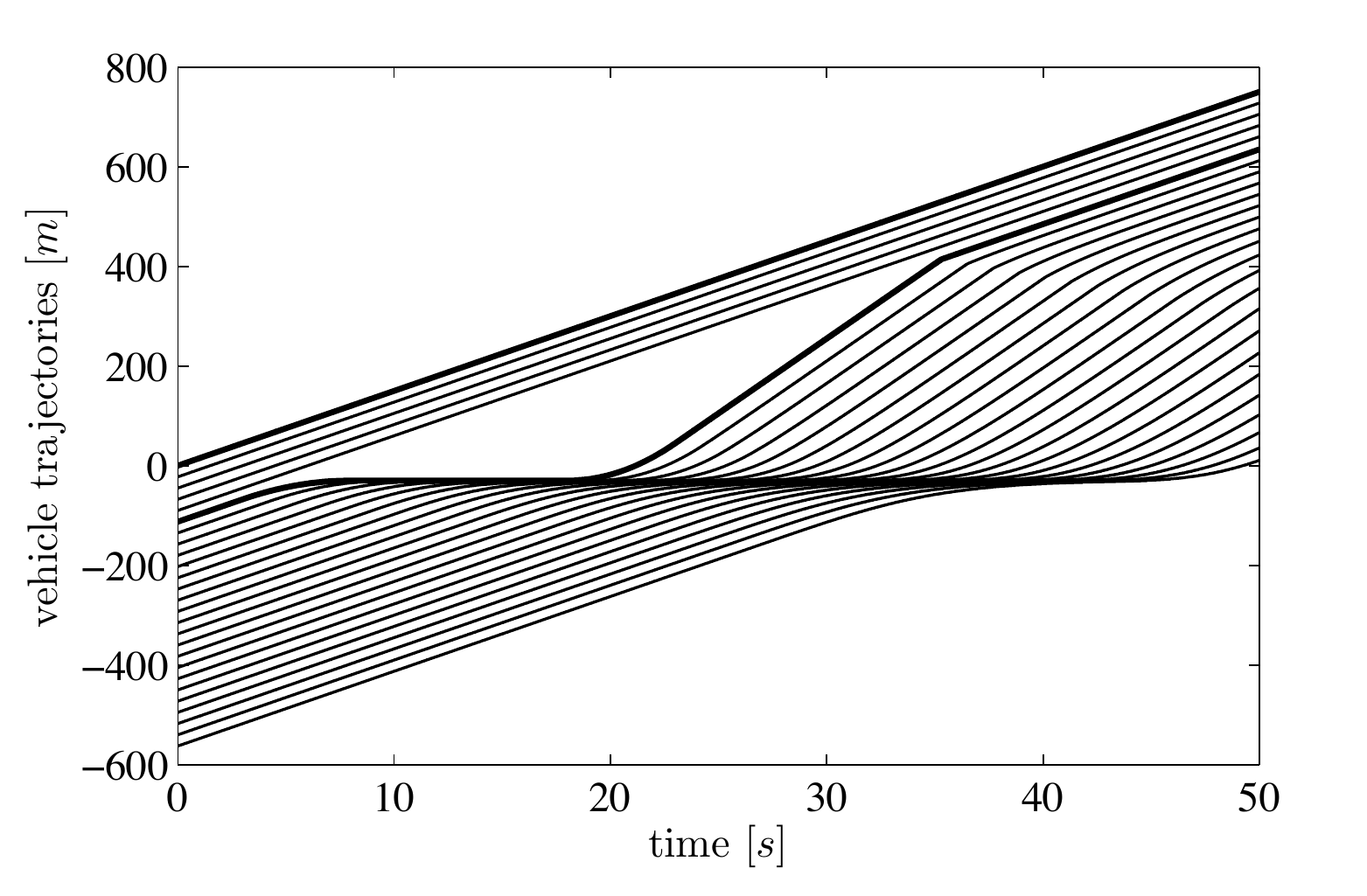}}
 \end{centering}
\caption{Separation and subsequent merging of a flow. The thick black lines correspond to the leading vehicles. The figure has been drawn for $c_0=30[m/s],~~c/c_0 =1.25$ and $\tau= 25.0$. }
\label{complicated}
\end{figure} 

For simplicity of the preceding discussion we have supposed that when there are two platoons the platoon in front remained unperturbed.  Clearly our discussion does not depend on this and both platoons could be undergoing perturbed motions.


\section{Concluding remarks.}

In this paper we have formulated a linearized theory of traffic flow and shown that it can qualitatively describe several disparate phenomena.  The linearization is carried out by focusing on solutions close to a steady uniform flow of a general class of equations of motion. We used the closed-form solution of a boundary-initial value problem to describe several phenomena.  Specifically, the illustrate the smooth variation of the velocity field in stop-and-go traffic, the discontinuous velocity field with shock waves in a traffic light problem, and discontinuous displacement fields in a problem where a single platoon of vehicles splits into two, and later merges back into one. 
Needless to say we are not arguing that traffic models should be linear.  Instead our point is that linearized models can often provide useful insight, which when needed may be followed-up with a numerical solution of the full nonlinear model.



\end{document}